# GHz fundamental mode-locking of a highly integrated Er-doped all-fiber ring laser

*Maolin Dai [1,2], Bowen Liu [1,2], Yifan Ma [1,2], Ruoao Yang [3], Zhigang Zhang [3], Sze Yun Set [1,2], and Shinji Yamashita [1,2]*

[1]*Department of Electrical Engineering and Information Systems, The University of Tokyo, Bunkyo-ku, Tokyo 113-8656, Japan*
[2]*Research Center for Advanced Science and Technology, The University of Tokyo, Meguro-ku, Tokyo 153-8904, Japan*
[3]*State Key Laboratory of Advanced Optical Communication Systems and Networks, School of Electronics, and Center for Quantum Information Technology, Peking University, Beijing 100871, China*

*maolin@cntp.t.u-tokyo.ac.jp*
*set@cntp.t.u-tokyo.ac.jp*
*syama@cntp.t.u-tokyo.ac.jp*

**Abstract:** High repetition rate ultrafast fiber lasers are important tools for both fundamental science and industry applications. However, achieving over GHz repetition rate in passively mode-locked fiber ring lasers is still challenging. Here, we demonstrate the first ring-cavity Er-doped fiber laser that achieves over GHz fundamental repetition rate by using an all-integration cavity design. In the proposed laser oscillator, all functions are integrated into one device, making it an ultra-compact laser cavity. The laser is mode-locked by carbon nanotubes (CNTs) film that is directly deposited on the pigtail active fiber connectors. The laser produces ultrafast optical pulses at 1562 nm, with a pulse width of 682 fs and a fundamental repetition rate of 1.028 GHz with improved performance. Stable and low-noise mode-locking is characterized by high signal-to-noise ratio (SNR) radiofrequency signal and low relative intensity noise (RIN). The proposed all-integration laser design may serve as a reference for compact fiber ring lasers using other mode-locking mechanisms or at diverse wavelengths.

1. Introduction

High repetition rate mode-locked fiber lasers are seen as the idea sources for optical arbitrary waveform generation [1], optical frequency comb-based metrology or spectroscopy [2-4], material processing [5] and so on. Methods such as active mode-locking [6, 7], passively harmonic mode-locking [8-10] or mode filtering [11, 12] have shown the ability to achieve an ultrahigh repetition rate over tens of gigahertz (GHz), however, in terms of pulse quality and operation stability, passively fundamental mode-locking is more preferred.

For high-repetition-rate mode-locking, significant progresses based on linear cavities have been achieved, combining with the real saturable absorbers (SAs) and high-reflection fiber mirrors [13-20]. For example, 5-GHz fundamental repetition rate was achieved in 2005, in an Er-Yb co-doped fiber laser mode-locked by carbon nanotubes (CNT) [13]. Recently, the record of repetition rate in linear cavities has been improved to 21 GHz by using a semiconductor saturable absorber mirror (SESAM) [19]. Compared to linear fiber lasers, ring fiber lasers are preferred owing to unidirectional laser operation with an intracavity isolator, which gets rid of the parasitic reflection as well as the spatial hole burning effect. Especially the former, in linear laser cavities, the coupling between waveguiding fiber and fiber mirror will highly influence the laser performance.

High repetition rate Yb-doped ring fiber lasers have been realized using nonlinear polarization evolution (NPE) mechanism [21-23]. However, compared to Yb-doped fibers, Er-doped fibers can provide much less gain for laser oscillation. Therefore, it is quite challenging to achieve high fundamental repetition rates in Er-doped fiber lasers with short fiber segments. In 2013, T. Yang et al. demonstrated 500 MHz fundamental repetition rate in an all-fiber ring laser with highly Er/Yb co-doped phosphate glass fiber and a polarization-sensitive integrated module [21]. In 2017, W. Du et al. demonstrated an NPE-mode-locked all-fiber ring laser using commercialized silica gain fiber, delivering a fundamental repetition rate of 384 MHz [24]. However, further increasing the repetition rate in NPE fiber ring lasers requires free-space components for polarization management, which sacrifices compactness and robustness [25]. As NPE mode-locking is powered by high intracavity peak power, therefore, extremely high pump power is required for those short-cavity lasers [25, 26]. Using real SAs can effectively mode-lock the high repetition rate lasers with self-starting using lower pump power, even using commercialized Er-doped silica fibers [18, 20]. J. W. Nicholson et al. demonstrated a 447-MHz mode-locked all-fiber ring laser using a CNT mode-locker, with 20-cm Er-doped fiber and 26-cm single-mode fiber [14]. Further cutting the SMF for a high repetition rate is quite challenging. So far, no mode-locked all-fiber ring lasers with GHz fundamental repetition rate have been reported.

Here, we demonstrate the first all-fiber ring laser with fundamental repetition rate exceeding GHz using a novel highly integrated fiber ring cavity design. In the proposed laser, all the necessary functions are integrated into a polarization-insensitive tap-isolator-wavelength division multiplexer (PI-TIWDM) with commercialized Er-doped fiber pigtails. The CNT-SA is directly deposited on the fiber connector, making it an ultra-compact fiber ring cavity with in-line mode-locker. The self-started fiber laser emits ultrafast optical pulses at 1562 nm, with a pulse width of 682 fs and a fundamental repetition rate of 1.028 GHz. The mode-locking stability is confirmed by the high signal-to-noise ratio (SNR) and low relative intensity noise (RIN).

## 2. Laser cavity configuration

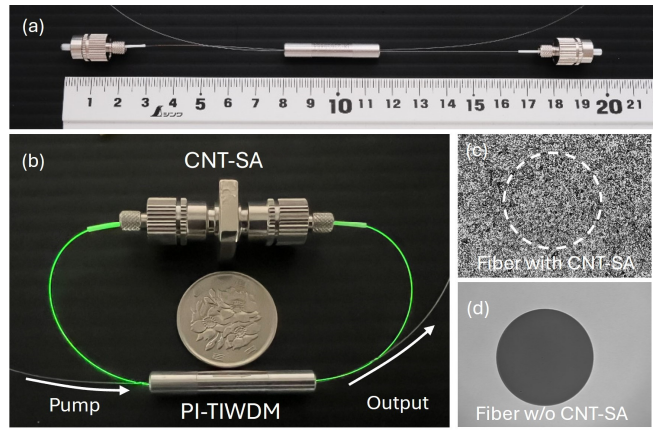

**Fig. 1.** Setup of the proposed ultrafast all-fiber ring laser with repetition rate exceeding GHz: (a) Open loop state, (b) close loop state. The PI-TIWDM integrates the functions of polarization-insensitive isolator, 10% output coupler and 980/1550 nm WDM. CNT-SA is sandwiched between two PC connectors. (c) Image of the PC connector with CNT-SA. (d) Image of the PC connector without CNT-SA.

The setup of the proposed all-fiber ring laser is shown in Fig. 1(a) and (b) with open loop state and close loop state, respectively. The laser cavity only consists of a polarization-insensitive tap/isolator/wavelength division multiplexer (PI-TIWDM) with pigtails of Er-doped fiber (EDF). A 976-nm laser diode (LD, 3SP Technologies) is used as the pump source for the laser system. The EDF we use is commercialized silica gain fiber (Liekki, 80-8/125), with a group velocity dispersion (GVD) of -20 $fs^2$/mm and a peak absorption of 80 dB/m at 1530 nm. Therefore, the laser is expected to operate in the soliton regime owing to the anomalous dispersion. The whole length of the compact laser cavity is 20 cm, with 3.5 cm length of the PI-TIWDM [Fig. 1(a)]. Considering both sides of the TIWDM should have about 1-cm fiber to couple, the total fiber length should be 18.5 cm. Therefore, cavity dispersion is estimated as -3700 $fs^2$. Both pigtails have physical contact (PC) connectors for CNT mode-locker integration. No polarization controllers are implemented in the cavity.

The CNT-SA is fabricated following the same methodology described in Ref. [20]. Figures 1(c) and (d) show the cross sections of the two facets of the two PC connectors, with and without CNT-SA deposition. The uniform CNT-SA film covers the fiber area, making a good light-matter interaction for initiating the mode-locking. As all the components are well in-line integrated, the laser cavity is ultra-compact with no passive fibers. Therefore, the laser cavity can fully make use of the cavity length for the maximum gain. To measure the saturable absorption of the CNT-SA, we fabricate the SA on the passive fiber patch cord under the same condition and use I-scan method to obtain the nonlinear transmission [27]. The result is shown in Fig. 2. With the peak intensity increases to ~200 MW/$cm^2$, the transmission after the sample is increased from 45.3% to 47.5%, yielding a modulation depth (MD) of 2.2%. The non-saturable loss is calculated to be 53.6% according to the fitting curve.

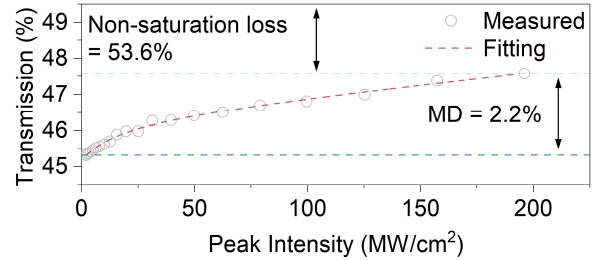

**Fig. 2.** Nonlinear transmission curve of the CNT-SA.

For laser output characterization, the optical spectrum is measured and analyzed by an optical spectrum analyzer (OSA, Ando, 6317B). The autocorrelation (AC) trace of the pulse is measured by an optical pulse analyzer (Southern Photonics, HR-150). The optical signal is transferred to electrical signal via a fast photodetector (New Focus, 1592), and then measured by an oscilloscope (Keysight, DSA91304A) and radiofrequency (RF) spectrum analyzer (RIGOL, RSA3045). The relative intensity noise (RIN) of the laser output is measured by the RIN analyzer (Thorlabs, PNA1).

## 3. Experimental results

The mode-locking is self-started when the pump power is set to ~222 mW. When turning down the pump power, the mode-locking threshold reaches ~200 mW owing to the pump hysteresis. Figure 3(a) shows the relationship between the output power and the injected pump power, as well as the operation status of the laser. Figure 3(b) shows the typical mode-locking evolution process when the pump power increases. At the highest stable single-pulse pump power of ~300 mW, the output power of the laser is measured as 1.65 mW. After laser mode-locked, further increasing the pump power will cause the broadening of optical spectrum as well as compression of pulse width under soliton area theorem [28].

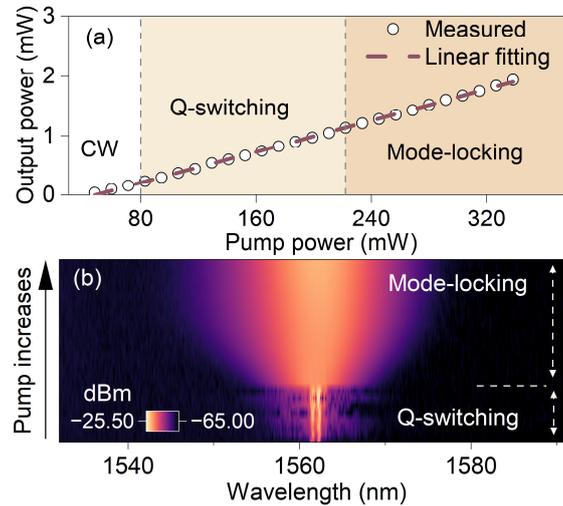

**Fig. 3.** (a) Power and (b) spectral evolution in the process of increasing pump power. When pump power reaches 222 mW, the laser enters the mode-locking regime with self-starting.

In the process of increasing pump power, the variations in center wavelength and 3-dB bandwidth are recorded as Fig. 4(a) and (b), respectively. The orange area represents the single-pulse mode-locking. With the pump power increasing, the laser transits from Q-switched to mode-locked with a spectrum broadening, which means the longitudinal modes are fixed. The 3-dB bandwidth of the mode-locked pulse increases from 4 nm to 6.8 nm with the pump power increased from 220 mW to 300 mW. With higher intracavity peak power, the pulse width should be decreased under the soliton area theorem [28]. Therefore, in the spectrum domain, the 3-dB bandwidth is continuously increased. Before laser mode-locked, the center wavelength randomly jumps in a small range owing to the mode competition, as no narrow filtering effect in the laser cavity. After mode-locked, the laser frequency is stabilized 1561.7 nm, and red-shifts to 1562 nm when reaches the maximum single-pulse pump power. The red shift of center wavelength can be attributed to the variation in EDF gain emission and the positive thermo-optic coefficient of the silica material.

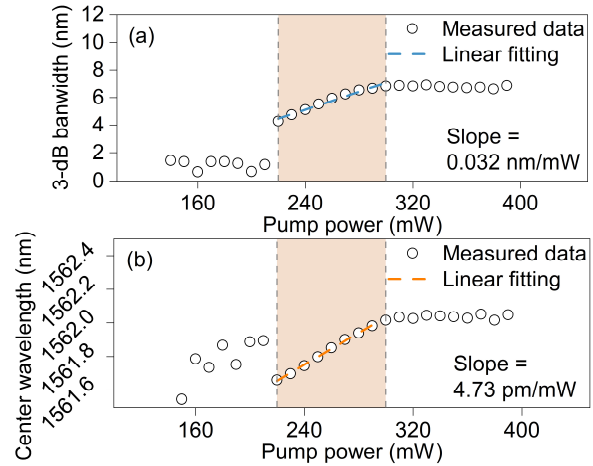

**Fig. 4.** Evolution of the (a) center wavelength and (b) 3-dB bandwidth with increasing pump power.

At the pump power of ~300 mW, the optical spectrum and the corresponding AC trace are measured, as shown in Fig. 5 (a) and (c) respectively. The results show the laser has a typical soliton-like spectrum with a center wavelength of 1562 nm and full-width at half-maximum (FWHM) of 6.8 nm. No obvious Kelly sidebands are observed, owing to the limited laser spectral width. In the temporal domain, the AC trace is well fitted by the $sech^2$ curve with a FWHM of 1.05 ps. Therefore, the laser has a pulse width of 682 fs (1050 fs/1.54). The time-bandwidth product (TBP) is calculated as 0.57, indicating the pulse is slightly chirped. The slight pulse broadening is mainly caused by the chromatic dispersion of the optical fiber outside cavity during the measurement. It's noted that, compared to our previously reported 783-MHz mode-locked laser, the laser reported in this work has a wider optical spectrum and a higher mode-locking threshold [20]. The reason can be attributed to the less dispersion of the cavity and less gain from the active fiber owing to the shorter fiber length. Correspondingly, the oscillogram of the laser pulse train in 20 ns span is shown in Fig. 5(b). The pulse train has a pulse interval of 0.98 ns, which corresponds to a repetition frequency of 1.02 GHz. There is no obvious variation in the intensity for each pulse, showing the good mode-locking stability. In a broader temporal window of 1 ms, the oscillogram shows no envelope with ~kHz frequency, indicating the laser is operating at continuous-wave (CW) mode-locking regime, without Q-switching component. Figures 5(d) and the inset show the RF spectra with resolution bandwidths (RBWs) of 100 Hz and 30 kHz, respectively. The spectrum shows that the laser has a fundamental repetition rate of 1.0282 GHz, which is the highest among all the ring fiber lasers demonstrated so far. The signal-to-noise ratio (SNR) is measured to be 80 dB, showing the robustness of the mode-locking. In the wide-range spectrum, we can see the four harmonics (0–4.5 GHz) of the repetition frequency with high SNR of over 40 dB even under a lower RBW. Considering the average output power of 1.67 mW and the pulse width of 682 fs, the laser has a pulse energy of 1.62 pJ and an output peak power (assuming $sech^2$) of 2.1 W.

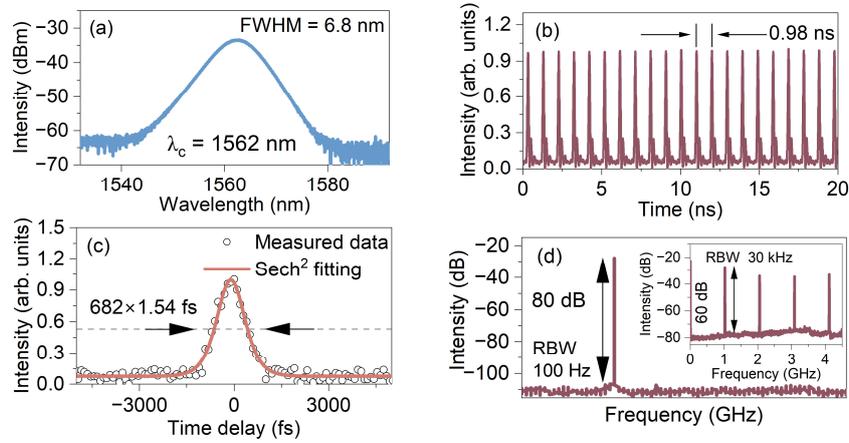

**Fig. 5.** Output performance of the proposed GHz all-fiber ring laser. (a) Pulse train in 20-ns window, (b) pulse train in 1-ms window, (c) RF spectrum with an RBW of 100 Hz and a span of 1 MHz, (d) RF spectrum with an RBW of 30 kHz and a span of 4.5 GHz.

To evaluate the noise performance of the laser, we characterize the RIN of the laser output after appropriate attenuation. Figure 6 shows the measurement results with 16 times average in the Fourier frequency offset from 1 MHz to 10 Hz. In the low frequency range of 100 – 10 Hz, the noise may contribute to mechanical and acoustic noise. Specifically, the small peak of 50 Hz corresponds to the frequency of alternating current. In the range of 100 kHz – 10 kHz, the RIN drops dramatically to near 138 dBc/Hz, which reaches the shot noise limit of 138.95 dBc/Hz. The shot noise $S(f)$ can be calculated by $S(f) = 2h\nu_c/P_{avg}$. Here, $h$ is the Planck's constant, $\nu_c$ is the center frequency of the laser output, and $P_{avg}$ is the average power injected into the photodiode. The $P_{avg}$ is attenuated to around 20 µW in our measurement, considering the saturation power of 55 µW of the photodiode (New Focus, 1811). The integrated root-mean-square (RMS) RIN is calculated as 0.028%, 0.045%, 0.047% and 0.049% respectively, from 1 MHz to 10 kHz, 1 kHz, 100 Hz and 10 Hz. The measured RIN performance of the proposed laser is comparable to that of the GHz fiber laser mode-locked by NPE [25].

Compared to our previous work of 783-MHz laser, the performance of laser in this work is significantly improved. Although the cavity is further shortened, the SNR of is improved from 70 dB (RBW 300 Hz) to 80 dB (RBW 100 Hz), showing the mode-locking is more stable [20]. Regarding the RIN performance, the GHz laser in this work reaches the shot noise limit above 300 kHz, whereas the 783-MHz laser exhibits an RIN-PSD consistently above -130 dB/Hz across the entire noise measurement range. The integrated RMS-RIN of the 783-MHz laser is 0.058% within the 1 MHz–100 Hz frequency range, while the corresponding integrated RMS-RIN of the laser in this work is 0.047%, demonstrating a significant improvement in noise performance. We attribute the stable mode-locking primarily to the use of standard FC/PC connectors instead of bare ferrules, which allows better fiber connection. The improved noise performance is not only associated with the stable mode-locking state but also results from the reduced dispersion induced by shorter intracavity fiber [29].

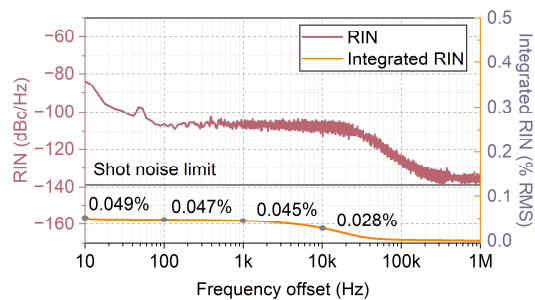

**Fig. 6.** Relative intensity noise performance of the proposed laser.

Table 1. Comparisons between mode-locked fiber ring lasers with high fundamental repetition rates.

| Refs | C.D. | F.R.R. | G.F. | P.W. | P.P. | P.E. | M.L.M. |
|---|---|---|---|---|---|---|---|
| [24] | All-fiber | 374 MHz | Liekki Er80-8/125 | 610 fs (uncompressed) | 650 mW | 39 pJ (unamplified) | NPE |
| [14] | All-fiber | 447 MHz | - | 270 fs | - | - | CNT |
| [21] | All-fiber | 500 MHz | Er/Yb Co-doped fiber | 250 fs | 650 mW | 7 pJ | NPE |
| [26] | Non-all-fiber | 517MHz | Liekki Er80-8/125 | 97 fs | 2 W | 180 pJ | NPE |
| [25] | Non-all-fiber | 1 GHz | Liekki Er80-8/125 | 81.8 fs | 3.2 W | 228 pJ | NPE |
| This work | All-fiber | 1 GHz | Liekki Er80-8/125 | 682 fs | 300 mW | 1.62 pJ | CNT |

C.D.: Cavity design; F.R.R.: Fundamental repetition rate; G.F.: Gain fiber; P.W.: Pulse width; P.P.: Pump power; P.E.: Output pulse energy; M.L.M: Mode-locking mechanism.

To compare the performance of the proposed high repetition rate ultra-compact all-fiber ring laser, Table. I lists the main works with fundamental repetition rates with ring cavity configurations. Among the listed works, we have achieved the highest repetition rate using commercialized Er-doped fiber with all-fiber cavity design. Compared to NPE lasers, the proposed laser has a lower pump power threshold as well as output power. The output power of our proposed laser can be further amplified, and the pulse width could be compressed by subsequent systems. Using real SAs enables the laser to have a better self-starting ability. In our experiment, we achieve the self-starting mode-locking repeatably when turning on the pump power. It's noted that no NPE lasers achieve repetition rates over 500 MHz with all-fiber configuration owing to the limited space for polarization management. Our laser cavity design that directly aligns the active fiber with WDM device may provide a possible way to solve this challenge combining with in-line polarization controller.

## 4. Conclusions

We have proposed and demonstrated a passively mode-locked, all Er-doped fiber ring laser with a fundamental repetition rate exceeding 1 GHz. By integrating all necessary functions into a single device, an ultracompact all-fiber ring laser that emits ultrafast femtosecond pulses at 1562 nm with a repetition rate of 1.028 GHz is achieved. The stable mode-locking is characterized by the high SNR of 80 dB and the low integrated RIN of 0.049% in the frequency range of [1 MHz–10 Hz]. For prospects, we estimate that the repetition rate can be further increased to approximately 2 GHz by reducing the size of the TIWDM and further shortening the fiber length. The CNT-SA has been proven in previous studies to effectively mode-lock lasers with ultrashort cavities, therefore, the primary challenges lie in the engineering techniques required to integrate all necessary functions into a more compact configuration. We believe the proposed laser holds potential for applications requiring high repetition rate optical pulses and provides guidance for high repetition rate fiber ring laser design.

**Funding.** This work was supported by the Japan Society for the Promotion of Science (JSPS) under Grant 23H00174 and Grant 22H00209.

**Data availability.** Data underlying the results presented in this paper are not publicly available at this time but may be obtained from the authors upon reasonable request.

**References**

1. S. T. Cundiff, and A. M. Weiner, "Optical arbitrary waveform generation," Nature Photonics **4**(11), 760-766 (2010).
2. T. Fortier, and E. Baumann, "20 years of developments in optical frequency comb technology and applications," Communications Physics **2**((2019).
3. N. Picqué, and T. W. Hänsch, "Frequency comb spectroscopy," Nature Photonics **13**(3), 146-157 (2019).
4. T. Voumard, J. Darvill, T. Wildi, M. Ludwig, C. Mohr, I. Hartl, and T. Herr, "1-GHz dual-comb spectrometer with high mutual coherence for fast and broadband measurements," Opt. Lett. **47**(6), 1379-1382 (2022).
5. H. Kalayciog, P. Elahi, O. Akcaalan, and F. O. Ilday, "High-Repetition-Rate Ultrafast Fiber Lasers for Material Processing," Ieee Journal of Selected Topics in Quantum Electronics **24**(3) (2018).
6. J. Qin, R. Dai, Y. Li, Y. Meng, Y. Xu, S. Zhu, and F. Wang, "20 GHz actively mode-locked thulium fiber laser," Opt. Express **26**(20), 25769-25777 (2018).

7. G. Yao, Z. G. Zhao, Z. J. Liu, X. B. Gao, and Z. H. Cong, "High repetition rate actively mode-locked Er:fiber laser with tunable pulse duration," Chinese Optics Letters **20**(7) (2022).
8. C. M. Wu, and N. K. Dutta, "High-repetition-rate optical pulse generation using a rational harmonic mode-locked fiber laser," IEEE J. Quantum Electron. **36**(2), 145-150 (2000).
9. C. S. Jun, J. H. Im, S. H. Yoo, S. Y. Choi, F. Rotermund, D.-I. Yeom, and B. Y. Kim, "Low noise GHz passive harmonic mode-locking of soliton fiber laser using evanescent wave interaction with carbon nanotubes," Opt. Express **19**(20), 19775-19780 (2011).
10. Z. Zhao, L. Jin, S. Y. Set, and S. Yamashita, "2.5 GHz harmonic mode locking from a femtosecond Yb-doped fiber laser with high fundamental repetition rate," Opt. Lett. **46**(15), 3621-3624 (2021).
11. W. Q. Wang, W. F. Zhang, S. T. Chu, B. E. Little, Q. H. Yang, L. R. Wang, X. H. Hu, L. Wang, G. X. Wang, Y. S. Wang, and W. Zhao, "Repetition Rate Multiplication Pulsed Laser Source Based on a Microring Resonator," Acs Photonics **4**(7), 1677-1683 (2017).
12. X. Cao, J. Zhou, Z. Cheng, S. Li, and Y. Feng, "GHz Figure-9 Er-Doped Optical Frequency Comb Based on Nested Fiber Ring Resonators," Laser & Photonics Reviews **17**(11), 2300537 (2023).
13. S. Yamashita, Y. Inoue, K. Hsu, T. Kotake, H. Yaguchi, T. Tanaka, M. Jablonski, and S. Y. Set, "5-GHz pulsed fiber Fabry-Perot laser mode-locked using carbon nanotubes," Ieee Photonics Technology Letters **17**(4), 750-752 (2005).
14. J. W. Nicholson, and D. J. Digiovanni, "High-Repetition-Frequency Low-Noise Fiber Ring Lasers Mode-Locked With Carbon Nanotubes," IEEE Photonics Technology Letters **20**(24), 2123-2125 (2008).
15. A. Martinez, and S. Yamashita, "Multi-gigahertz repetition rate passively modelocked fiber lasers using carbon nanotubes," Opt. Express **19**(7), 6155-6163 (2011).
16. A. Martinez, and S. Yamashita, "10 GHz fundamental mode fiber laser using a graphene saturable absorber," Appl. Phys. Lett. **101**(4) (2012).
17. X. Chen, W. Lin, W. Wang, X. Guan, X. Wen, T. Qiao, X. Wei, and Z. Yang, "High-power femtosecond all-fiber laser system at 1.5 μm with a fundamental repetition rate of 4.9 GHz," Opt. Lett. **46**(8), 1872-1875 (2021).
18. X. Gao, Z. Zhao, Z. Cong, G. Gao, A. Zhang, H. Guo, G. Yao, and Z. Liu, "Stable 5-GHz fundamental repetition rate passively SESAM mode-locked Er-doped silica fiber lasers," Opt. Express **29**(6), 9021-9029 (2021).
19. X. Chen, W. Lin, X. Hu, W. Wang, Z. Liang, L. Ling, Y. Yang, Y. Guo, T. Liu, D. Chen, X. Wei, and Z. Yang, "Dynamic gain driven mode-locking in GHz fiber laser," Light: Science & Applications **13**(1), 265 (2024).
20. M. L. Dai, B. W. Liu, Y. F. Ma, T. Shirahata, R. A. Yang, Z. G. Zhang, S. Y. Set, and S. Yamashita, "783 MHz fundamental repetition rate all-fiber ring laser mode-locked by carbon nanotubes," Applied Physics Express **17**(6) (2024).
21. T. Yang, H. C. Huang, X. Z. Yuan, X. M. Wei, X. He, S. P. Mo, H. Q. Deng, and Z. M. Yang, "A Compact 500 MHz Femtosecond All-Fiber Ring Laser," Applied Physics Express **6**(5) (2013).
22. C. Li, Y. Ma, X. Gao, F. Niu, T. Jiang, A. Wang, and Z. Zhang, "1 GHz repetition rate femtosecond Yb: fiber laser for direct generation of carrier-envelope offset frequency," Appl. Opt. **54**(28), 8350-8353 (2015).
23. R. Yang, M. Zhao, X. Jin, Q. Li, Z. Chen, A. Wang, and Z. Zhang, "Attosecond timing jitter from high repetition rate femtosecond "solid-state fiber lasers"," Optica **9**(8), 874-877 (2022).
24. W. X. Du, H. D. Xia, H. P. Li, C. Liu, P. H. Wang, and Y. Liu, "High-repetition-rate all-fiber femtosecond laser with an optical integrated component," Appl. Opt. **56**(9), 2504-2509 (2017).
25. C. Zhendong, R. Yang, Y. Tang, D. Pan, J. Cao, M. Zhang, Z. Zhang, and J. Chen, "GHz fundamental repetition rate femtosecond Er: fiber laser mode locked by nonlinear polarization evolution,"  (2024).
26. J. Zhang, Z. Kong, Y. Liu, A. Wang, and Z. Zhang, "Compact 517 MHz soliton mode-locked Er-doped fiber ring laser," Photon. Res. **4**(1), 27-29 (2016).
27. M. Dai, B. Liu, G. Ye, T. Shirahata, Y. Ma, N. Yamaguchi, S. Y. Set, and S. Yamashita, "Pump-driven wavelength switching in an all-polarization-maintaining mode-locked fiber laser incorporating a CNT/PDMS saturable absorber," Optics & Laser Technology **176**(111002 (2024).
28. G. P. Agrawal, "Chapter 5 - Optical Solitons," in *Nonlinear Fiber Optics (Fifth Edition)*, G. Agrawal, ed. (Academic Press, 2013), pp. 129-191.
29. J. Kim, and Y. Song, "Ultralow-noise mode-locked fiber lasers and frequency combs: principles, status, and applications," Advances in Optics and Photonics **8**(3), 465-540 (2016).